\documentclass[journal=nalefd,manuscript=letter]{achemso}

\usepackage{chemformula} 
\usepackage[T1]{fontenc} 
\usepackage{dcolumn}
\usepackage{bm}
\usepackage{amsmath,amssymb}
\usepackage{times}
\usepackage{float}
\usepackage{graphicx, epstopdf}
\usepackage{color}
\usepackage{setspace}
\usepackage{algorithm2e}
\usepackage{notoccite}
\usepackage{siunitx}
\usepackage{natbib}
\usepackage{epsfig,amsfonts}
\usepackage[caption=false]{subfig}
\usepackage{upgreek}
\usepackage[noabbrev]{cleveref}
\usepackage{scalerel}
\usepackage[export]{adjustbox}

\author{Paolo Andrich}
\affiliation[University of Chicago]{Institute for Molecular Engineering, University of Chicago, Chicago, IL 60637, USA}
\altaffiliation{These two authors contributed equally}

\author{Jiajing Li}
\affiliation[University of Chicago]{Institute for Molecular Engineering, University of Chicago, Chicago, IL 60637, USA}
\altaffiliation{These two authors contributed equally}

\author{Xiaoying Liu}
\affiliation[University of Chicago]{Institute for Molecular Engineering, University of Chicago, Chicago, IL 60637, USA}

\author{F. Joseph Heremans}
\affiliation[Argonne National Lab]{Institute for Molecular Engineering and Materials Science Division, Argonne National Lab, Argonne, IL 60439, USA}
\alsoaffiliation[University of Chicago]{Institute for Molecular Engineering, University of Chicago, Chicago, IL 60637, USA}

\author{Paul F. Nealey}
\affiliation[University of Chicago]{Institute for Molecular Engineering, University of Chicago, Chicago, IL 60637, USA}
\alsoaffiliation[Argonne National Lab]{Institute for Molecular Engineering and Materials Science Division, Argonne National Lab, Argonne, IL 60439, USA}

\author{David D. Awschalom}
\affiliation[University of Chicago]{Institute for Molecular Engineering, University of Chicago, Chicago, IL 60637, USA}
\alsoaffiliation[Argonne National Lab]{Institute for Molecular Engineering and Materials Science Division, Argonne National Lab, Argonne, IL 60439, USA}
\email{awsch@uchicago.edu}

\title[An \textsf{achemso} demo]   {Microscale resolution thermal mapping using a flexible platform of patterned quantum sensors}

\begin{document}







\begin{abstract}
  Temperature sensors with micro- and nanoscale spatial resolution have long been explored for their potential to investigate the details of physical systems at an unprecedented scale. In particular, the rapid miniaturization of transistor technology, with the associated steep boost in power density, calls for sensors that accurately monitor heating distributions. Here, we report on a simple and scalable fabrication approach, based on directed self-assembly and transfer printing techniques, to construct arrays of nanodiamonds containing temperature sensitive fluorescent spin defects. The nanoparticles are embedded within a low thermal conductivity matrix that allows for repeated use on a wide range of systems with minimal spurious effects. Additionally, we demonstrate access to a wide spectrum of array parameters ranging from sparser single particle arrays to denser devices with \SI{\sim 100}{\%} yield and stronger photoluminescence signal, ideal for temperature measurements. With these we experimentally reconstruct the temperature map of an operating coplanar waveguide to confirm the accuracy of these platforms. 
\end{abstract}

\section{Introduction}
\vspace{-0.3cm}

Recent years have seen dramatic advances in the field of nanotechnology, which pushes the boundaries of the engineering and control of matter at an unprecedented scale. In particular, efforts towards the miniaturization of electronic devices have 
driven silicon transistor technology towards its projected scaling limit of \SI{\sim 5}{\nano\meter} gate lengths\cite{Mamaluy2015}, and have spurred the emergence of alternative paradigms that could challenge that limit\cite{Franklin2015,Blanco2015,Perrin2015,Desai2016}. This trend has also created the need for sensing techniques that could enable the investigation of devices at a micro- and nanoscale level. The miniaturization of integrated circuits and the simultaneous increase in interconnects have  indeed resulted in a steep power density rise in these devices, and the resulting temperature stress on these electronic components could negatively impact their performance and reliability. For this reason, monitoring the temperature of these systems with high spatial resolution is of great importance to individuate possible hot spots and fail prone regions. 

Currently available sensing techniques for ambient conditions and high spatial resolution operation include liquid crystal thermography (LCT)\cite{Zharkova2002}, fluorescent microthermography (FMT)\cite{Barton1996}, scanning thermal microscopy (SThM)\cite{Cahill2001}, and thermoreflectance microscopy (TRM)\cite{Ryu2015}. While these approaches have prompted important advancements in the field\cite{Liu2007}, they also experience some crucial limitations. LCT and FMT require complex specimen preparations and experimental setups\cite{Fernicola2000}, and the employed sensing coating can be a source of spurious heat capacity and imprecision due to lack of flatness. SThM techniques are suitable for thermal mapping of relatively small areas due to the typically limited scanning range, and the sample's surface roughness cannot exceed a few micrometers. Additionally, the complexity of the probe design limits the scalability of these systems and can introduce large uncertainties due to the thermalization of the sample through other parts of the scanning device\cite{Majumdar}, especially in the presence of sharp scanning tips. While TRM has the advantage of being a completely contactless technique, it requires a sample-by-sample extensive calibration process, and the interpretation of the measurements is complex in the presence of textured samples that affect the probe reflection at the interface. 

To address these limitations, we develop a platform that provides both an optimal thermal contact between the sensors and the samples and minimal unwanted effects on the real temperature distribution. Additionally, our architecture requires only a single pre-calibration and can be repeatedly used to investigate a wide range of systems. 
The sensitivity is provided by the response of the spin levels of the nitrogen-vacancy (NV) center in diamond to variation in the local environmental temperature\cite{Toyli2013}. These defects can be embedded within nanoparticles (nanodiamonds, NDs) and the spin levels are optically addressable, making them ideal for remote, contactless operation. The positioning of the NDs is achieved using a directed assembly chemical patterning technique\cite{Xiaoying2015,Xiaoying2016} optimized for the NDs surface chemistry, which allows for the deterministic placement of sensors on a silicon sacrificial substrate. Finally, a transfer printing technique is employed to place the nanoparticles onto the surface of a polydimethylsiloxane (PDMS) layer. The result is an arrangement of thermal sensors with controllable spacing and properties encapsulated within a transparent and portable matrix with low thermal conductivity, which can be easily applied to the sample of interest. Remarkably, the use of commercially available NDs and established patterning techniques make this system suitable for large-area detection. 

In this letter, we present the details of the fabrication process and investigate the parameters that influence the sensor platform properties. Noticeably, we show that not only is it possible to create dense arrays of nanoparticles optimal for temperature measurements, but also sparse arrays with single isolated NDs for other nanosensing and quantum applications, as we have recently demonstrated\cite{Andrich2017}. These applications take advantage of the NV centers' sensitivity to magnetic\cite{Balasubramanian2008} and electric fields\cite{Dolde2011}, and their implementation is greatly simplified when the sensors are comprised of a single monocrystalline particle\cite{Horowitz2012}. Finally, we illustrate the capabilities of the sensors by mapping with micron-scale resolution the temperature distribution of a gold coplanar waveguide carrying a microwave signal. By comparing the experimental results with the simulated heat map we demonstrate that the sensing platform accurately measures hot regions in the microelectronic circuit. The simulation also confirms that the PDMS layer does not have a significant effect on the measured thermal profile.

\section{Fabrication of ordered, transferable nanodiamond arrays}

\begin{figure}[!t]
	\centering
	\subfloat[][]{\label{Workflow}\includegraphics[scale=0.45]{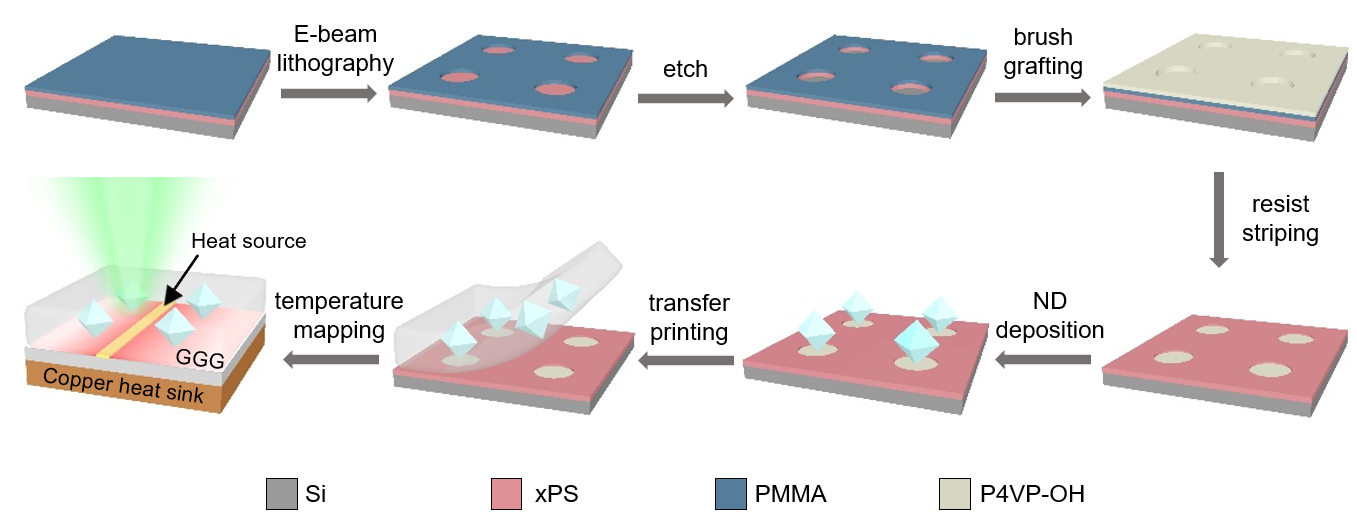}}\\
	\subfloat[][]{\label{SEMDiffSizes}\includegraphics[]{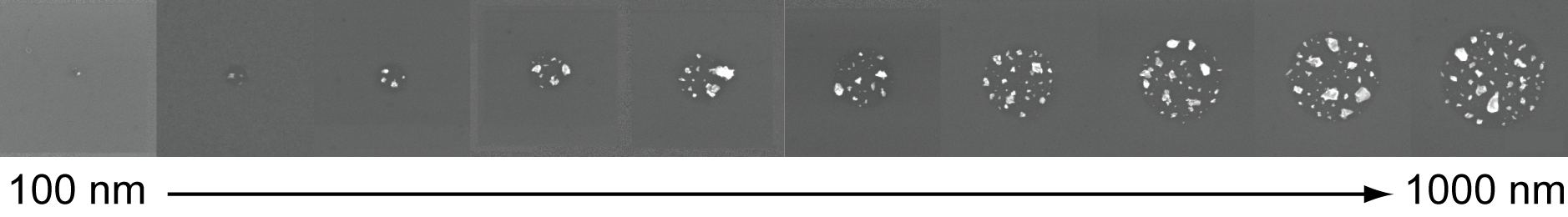}}
	\caption{(a) Fabrication process for the realization of arrays of nanodiamonds embedded within a transparent, flexible and portable matrix. First, crosslinked PS and PMMA resist are applied to a silicon substrate. Electron beam lithography, followed by resist development and a plasma ashing step are used to create the desired patterns within the PS layer. The sample is then functionalized with P4VP-OH, before removing the residual resist. Nanodiamonds from a drop casted solution bind preferentially at the remaining functionalized region and can finally be transferred into a PDMS layer. The schematic of the temperature measurements developed in this work is depicted as the last step of the fabrication workflow. The PDMS layer is positioned on top of a gadolinium-gallium-garnet substrate with a microwave antenna patterned on top, which acts as a heat source. A confocal microscopy setup with a 532 nm laser excitation is used to address the NV centers in the diamond nanoparticles. (b) SEM images of a series of polymer functionalized areas on a silicon substrate, coated with diamond nanoparticles. The diameter of the patterned circles varies from \SIrange{100}{1000}{\nano\meter} in steps of \SI{100}{\nano\meter}.}
\end{figure}

The fabrication workflow is schematically shown in \Cref{Workflow} and begins with the patterning of the nanodiamond arrays on a silicon chip (Si $<$100$>$, N/Phos, WRS Materials). The substrate is first spin coated with \SI{0.5}{wt\%} cross-linkable polystyrene (PS) in toluene and annealed at \SI{190}{\celsius} under vacuum for \SI{24}{} hours to drive the crosslinking reaction. The PS is synthesized as described in \cite{Han2008} and contains \SI{4}{\text{\%}} glycidyl methacrylate as a crosslinking agent. Spin-coating and curing of poly(methyl methacrylate) (PMMA950, \SI{4}{wt\%} in chlorobenzene, MicroChem Inc.) resist is followed by the electron beam exposure of the desired patterns. After the resist development with a mixture of 4-methyl-2-pentanone and 2-propanol (1:3 in volumetric ratio), we use an oxygen plasma ashing process (\SI{20}{sccm} O$_{2}$, \SI{20}{\watt}, \SI{30}{\second}) to remove the residual resist and the crosslinked PS (xPS) in the unprotected regions. The exposed silicon regions are functionalized with hydroxyl-terminated poly(4-vinylpyridine) (P4VP-OH, Polymer Source, Inc.) by spin coating from a \SI{4}{wt\%} solution in N,N-dimethylformamide (DMF), followed by annealing at \SI{210}{\celsius} for \SI{5}{\minute} in a nitrogen atmosphere to evaporate the solvent and promote the bonding of the P4VP-OH with the substrate. The remaining resist and excess P4VP-OH are then removed by sonication in 1-methyl-2-pyrrolidinone (NMP) (\SI{3}{\minute}, 2 cycles) and chlorobenzene (\SI{3}{\minute}, 1 cycle). A \SI{100}{\micro\liter} drop of ND suspension (\SI{2}{\micro\liter} of Adamas Technology, NV-ND-100nm in \SI{1.5}{\milli\liter} deionized (DI) water) is deposited on the substrate, which is placed on an elevated post inside a sealed glass jar. The jar also contains \SI{1}{\milli\liter} of water at the bottom to maintain a constant water vapor pressure and prevent the ND solution from evaporating. As the nanoparticles have a negative zeta potential ($\zeta$ \nobreakspace = \nobreakspace \SI{-35}{\milli\volt} in DI water), they are electrostatically attracted to the P4VP-OH polymers, which are weakly positively charged in water due to protonation, and thus the NDs bind at their sites. After \SI{40}{\minute}, the substrate is rinsed thoroughly with DI water and blown dry with nitrogen. The PS provides minimal adhesion of the NDs to the substrate resulting in a good selectivity of the regions patterned with P4VP-OH. 

Finally, the ND arrays are transferred onto the polymer matrix. First, PDMS base and curing agent (Sylgard 184, Dow Corning Corp.) are mixed together in a 10:1 mass ratio. The mixture is degassed in a vacuum chamber for one hour before being poured on top of the ND arrays. The sample is then cured at \SI{60}{\celsius} for \SI{12}{} hours, during which the PDMS hardens and becomes an elastic solid material. The sample is allowed to cool to room temperature before the thin PDMS layer, and with it a portion of the NDs, is removed from the substrate and transferred onto a clean silicon chip for characterization. 

\section{Characterization of the arrays of sensors}

\begin{figure}
	\centering
	\subfloat[][]{\label{NDPerArea}\includegraphics[height=6cm]{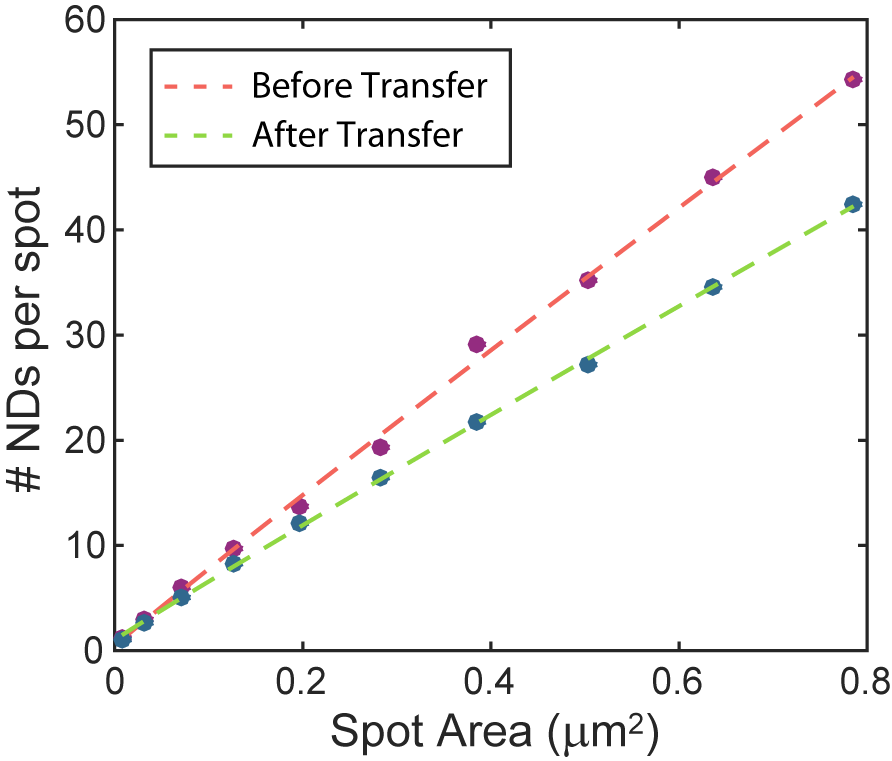}} 
	\subfloat[][]{\label{NDOccur}\includegraphics[height=6cm]{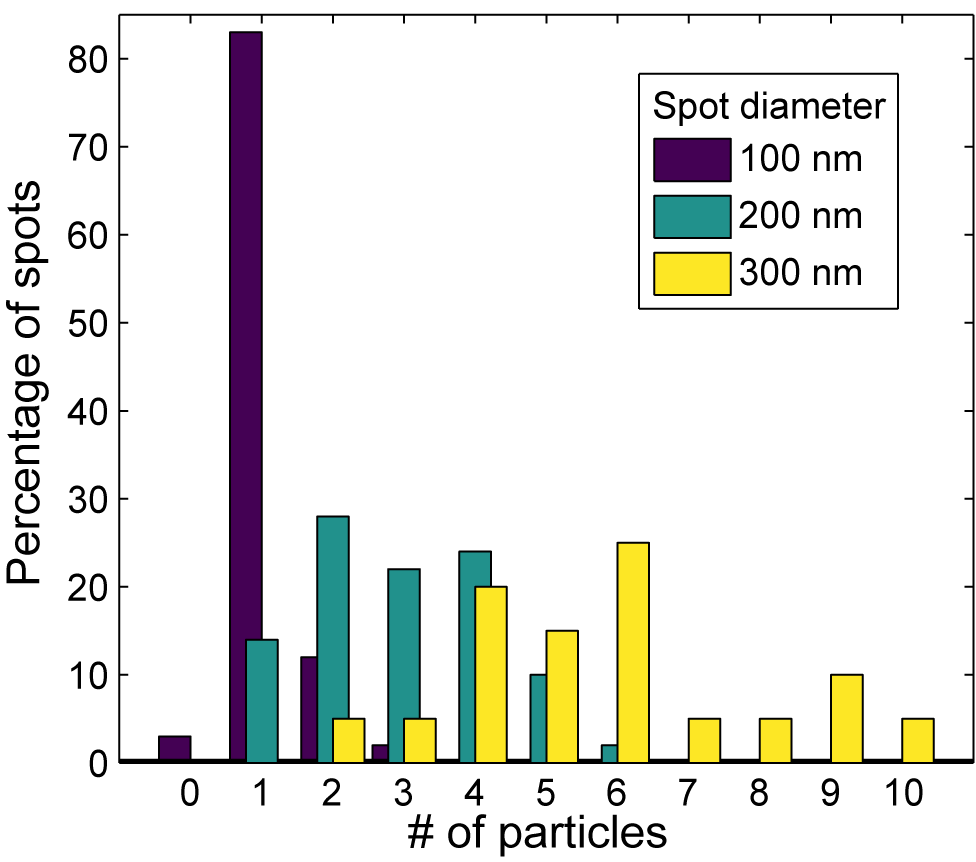}}
	\caption[Characterization of the number of nanodiamonds per patterned spot as a function of the spot diameter]{(a) Number of NDs per patterned spot as a function of the spot area $A$ before and after the PDMS transfer procedure. We fit the data to $y = cA^{\upalpha}$, where $c$ is a free proportional factor and $\upalpha$ assumes the value of 0.94 and 0.89 before and after the transfer respectively. (b) Histogram representation of the number of nanoparticles per spot for patterned circles with \SI{100}{}, \SI{200}{}, and \SI{300}{\nano\meter} diameters before the transfer procedure. The vertical axis shows the percentage of analyzed spots showing a certain number of bound NDs (horizontal axis).}
	\label{SEMStat}
\end{figure}

We characterize our fabrication process by investigating the effect of the patterning parameters on the properties of the ND arrays created on the silicon substrates. As detailed later, this analysis is important for understanding the fabrication of the PDMS bound arrays with specific nanoparticle densities.

In particular, we pattern square arrays of circular apertures with different diameters and spacings and study the resulting ND coverage yield and the number of nanoparticles per area of patterned substrate. 
In \Cref{SEMDiffSizes}, we show a series of SEM images collected from patterned circles with diameters ranging from \SIrange{100}{1000}{\nano\meter}, which reveal a direct relationship between the number of nanoparticles per spot and the spot size. Analyzing tens of patterned spots, we obtain a more quantitative description (\Cref{SEMStat}, purple data points) and show that the number of bound nanoparticles increases roughly linearly with the surface area of the patterned region (\Cref{NDPerArea}). No significant variation in the number of particles is observed when the pitch of the patterned arrays is changed from \SI{2}{\micro\meter} to \SI{10}{\micro\meter} (see for example Fig. S2). Interestingly, we show that by using \SI{100}{\nano\meter} diameter spots we can obtain arrays comprised of more than \SI{80}{\%} of single isolated NDs (\Cref{NDOccur}). We note that this result implies that this patterning technique can be used to create permanent arrangements of solid state qubits for sensing or quantum information purposes on any sample compatible with the fabrication process. Other approaches were recently used to position nanodiamonds on permanent substrates\cite{Kianinia2016,Heffernan2017}, although the control on the number on nanoparticles per spot was not demonstrated, and near \SI{100}{\%} yield and selectivity cannot be achieved simultaneously. Our fabrication process therefore also provides an important advance in the creation of stable ND arrays for quantum applications\cite{Schietinger2009}.

\Cref{NDOccur} also highlights the presence of a greater variability in the number of nanoparticles per spot for larger spot sizes, which we associate with the wide distribution of the nanoparticle size (\SIrange{\sim 50}{300}{\nano\meter}). While a \SI{100}{\nano\meter} patterned circle tends to be mostly covered by a single ND, independently of its size, the maximum coverage for larger circles can be reached with different numbers of nanoparticles. This variability could be addressed using a ND suspension with a narrower size distribution as obtained, for instance, through size separation in solution by centrifugation. 

We then proceed to characterize the efficiency of transfer of the nanoparticles into the PDMS layer by first conducting an SEM analysis of a series of patterned spots on the silicon substrate, subsequent to the transfer printing process, to estimate the number of nanoparticles that were removed. The result of these measurements is shown in \Cref{NDPerArea} (blue data points).
From these measurements we conclude that the number of NDs that are removed from the substrate increases with the area $A$ of the patterned circles. By fitting the data to  $y = cA^{\upalpha}$, where $c$ is a free proportionality factor, we obtain for $\upalpha$ the value of \SI{0.89}{} $\pm$ \SI{0.034}{}, compared with \SI{0.94}{} $\pm$ \SI{0.052}{} for the data collected before the transfer process. This implies that the transfer efficiency moderately depends on the size of the patterned areas, with the larger ensembles of NDs being transferred more easily. We attribute this result to the presence of an inverse relationship between the density of NDs per spot and the spot size (see Supplementary Fig.S3), which guarantees a better spreading of the PDMS between the nanoparticles with a consequent larger transfer efficiency. We note that the average size of the nanoparticles on the silicon substrate is not affected by the transfer process (Supplementary Fig.S3) suggesting that the transfer efficiency is independent of this parameter.  

\begin{figure}[!t]
	\centering
	\subfloat[][]{\label{NDArrayPL}\includegraphics[scale=0.9]{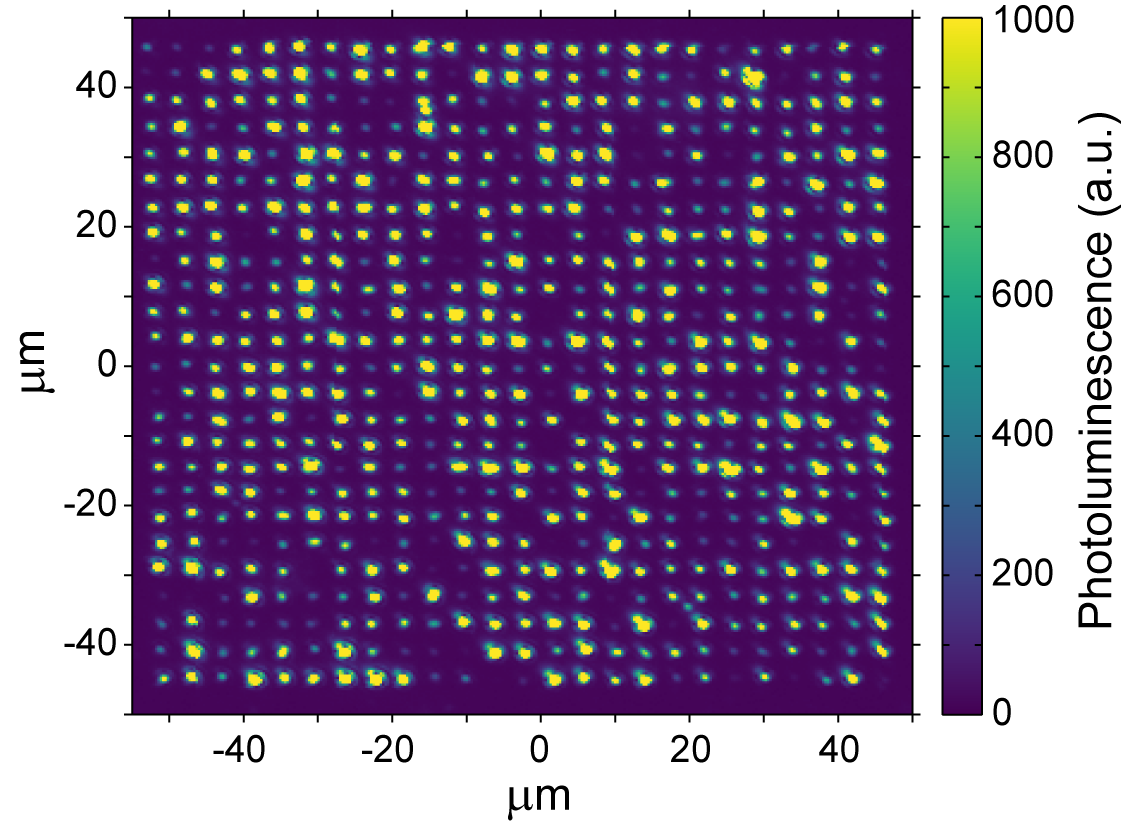}} 
	\subfloat[][]{\label{Coverage}\includegraphics[scale=0.9]{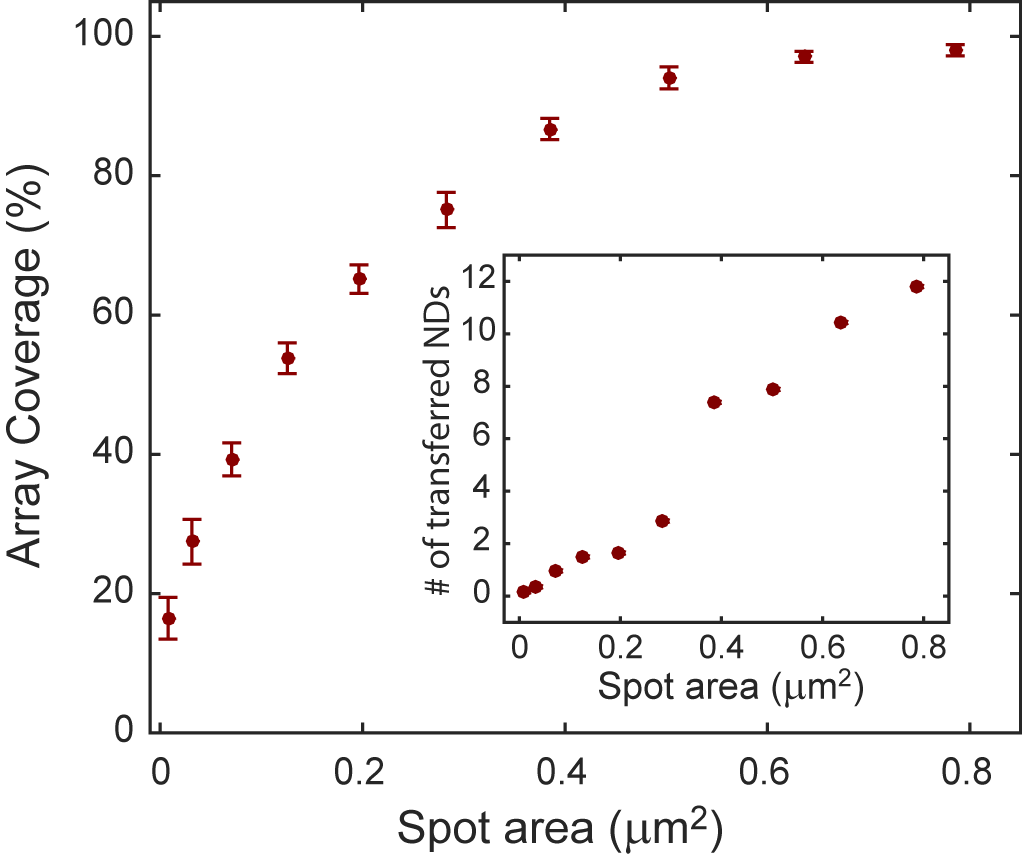}}
	\caption[Photoluminescence characterization of ND arrays in polydimethylsiloxane.]{(a) Two-dimensional PL scan of an array of NDs embedded within a PDMS matrix obtained using a 25 by 25 array of \SI{1000}{\nano\meter} diameter spots with \SI{4}{\micro\meter} spacing. We impose an upper bound to the value of the plotted PL to enable the visibility of most of the nanoparticles. (b) Percentage of the array sites with at least one nanoparticle with detectable PL signal as a function of the patterned spot area. We measure between 9 and 11 arrays per data point to achieve a sound statistical analysis. In the inset we report the expected number of nanoparticle per site in the PDMS matrix as calculated from the data in \Cref{NDPerArea}.}
	\label{NDArrayOpt}
\end{figure}

In the inset of \Cref{Coverage} we show the average number of nanoparticles that were removed from the substrate as a function of the patterned spot area as obtained from the data in \Cref{NDPerArea}. This allows us to estimate the number of NDs per spot in the transferred arrays. Nevertheless, the uncertainty on these figures (not shown in the plot for clarity) is comparable with their values, suggesting the presence of a wide spot-to-spot variability.

To gain further insight into the properties of the transferred ND arrays we characterize their photoluminescence (PL) by collecting two dimensional maps of the signal using a custom-built confocal microscopy setup. During these measurements the PDMS sample is placed on a bare silicon substrate with the nanodiamonds in contact with the silicon surface, and the PL is collected through the PDMS layer. In \Cref{NDArrayPL} we present a typical PL map of a ND array obtained from a sample patterned with \SI{1000}{\nano\meter} diameter spots and \SI{4}{\micro\meter} spacing. As expected, the PL signal shows wide variations between different spots. For each patterned spot size we collect data on a series of ten arrays to investigate their quality in terms of the coverage percentage, defined as the percentage of array sites that result in at least one nanodiamond transferred to the PDMS layer. The result of these measurements is reported in \Cref{Coverage}. The array coverage progressively increases with the area of the patterned spots and saturates close to the condition of full coverage (\SI{98.0}{} $\pm$ \SI{0.8}{\%}) for the largest spots. Comparing these results with the number of expected transferred nanoparticles we see that by using spots with \SI{0.2}{\micro\meter^{\scaleto{2}{5pt}}} area we can obtain arrays with \SI{\sim 65}{} $\pm$ \SI{2}{\%} coverage and $<$ 2 NDs per spot. This condition could be valuable for applications that require the use of single nanoparticles in order to simplify the interpretation of the data or optimize the PL contrast. On the other hand, temperature sensing applications at zero field can take advantage of arrays with higher ND density and a larger PL signal. 

\section{Temperature sensing with PDMS-nanodiamonds arrays}

Finally, we highlight the sensing capabilities of the PDMS-nanodiamond array system by demonstrating its potential as a thermal mapping technique. In particular, as a proof of principle, we image the spatial temperature profile generated by a coplanar waveguide (CPW) antenna that we concurrently employ to manipulate the NV centers' spin. 

\begin{figure}[!t]
	\centering
	\subfloat[][]{\label{TempMap}\includegraphics[scale=0.85]{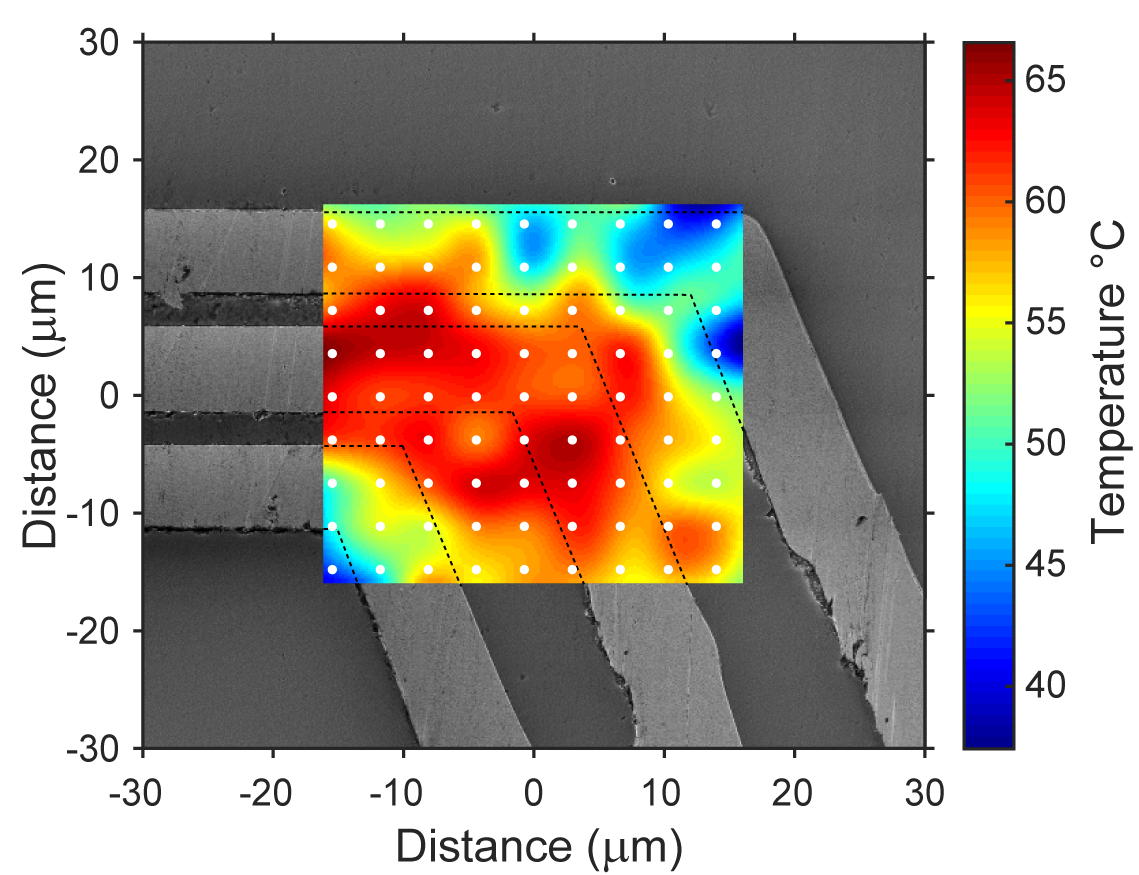}} 
	\subfloat[][]{\label{TempMapSim}\includegraphics[scale=0.85]{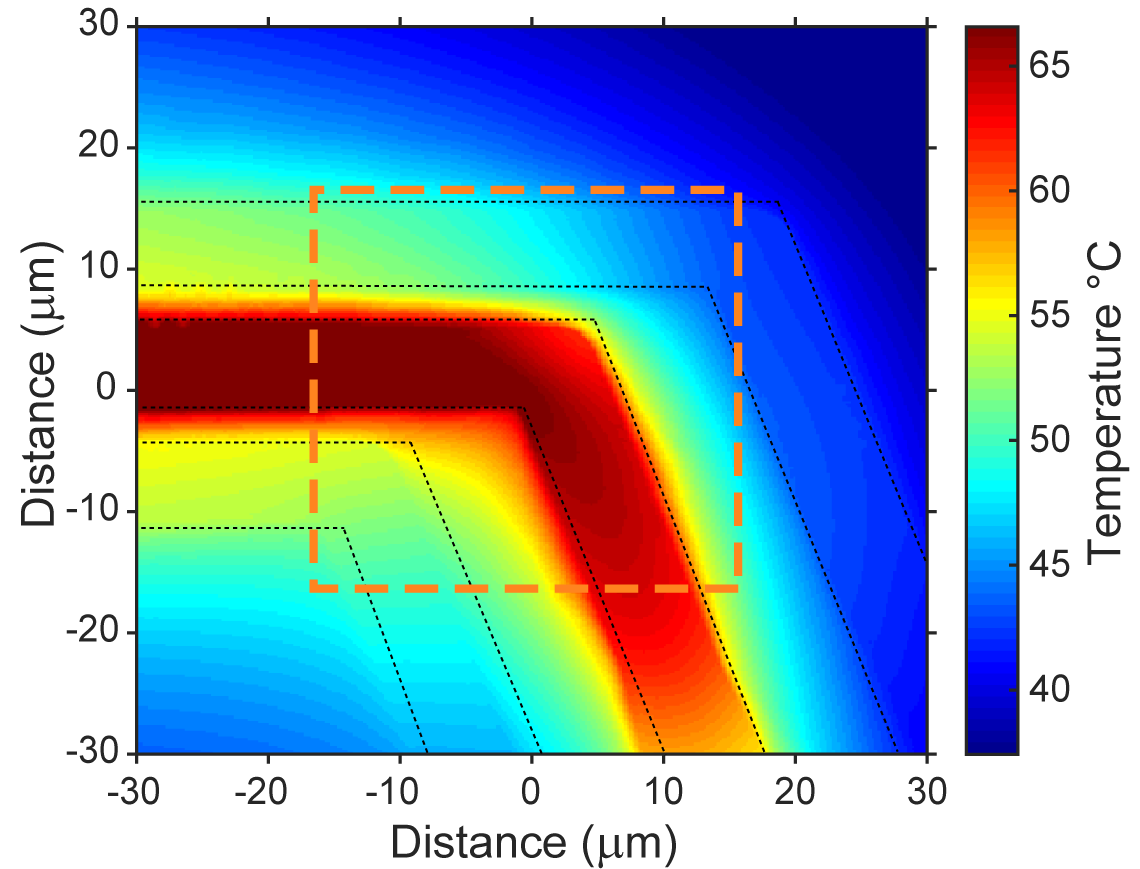}}
	\caption[Temperature map of a coplanar waveguide antenna obtained with an array of nanodiamonds in a PDMS matrix]{(a) Temperature map of a CPW obtained from the interpolated data collected on a 9 by 9 array of NDs (real positions indicated by the white dots) by detecting the shift in the zero-field splitting of ensembles of NV centers. The thermal image is superimposed on the SEM image of the CPW to identify the sample's features. (b) Simulation of the CPW's temperature map showing qualitative agreement with the data. The dashed box highlights the area that was investigated in the measurements shown in (a).}
	\label{TempMeas}
\end{figure}

We position a ND array obtained with \SI{1000}{\nano\meter} patterned spot diameter and \SI{4}{\micro\meter} spacing in contact with a CPW patterned through electron-beam lithography on a gadolinium gallium garnet (GGG) insulating substrate (see \Cref{Workflow} for the schematic of the sample) and explore a 9 by 9 particle array subset that is located at a sharp bend in the antenna (see \Cref{TempMeas}), which we expect to be affected by the largest heating effect. To perform the sensing measurements we use a modified continuous-wave optically detected magnetic resonance (ODMR) scheme shown in Supplementary Fig.S4. The microwave input is alternated between a \SI{100}{\micro\second}, \SI{0.4}{\watt} pulse off-resonant with respect to the NV centers' spin transitions (\SI{2.6}{\giga\hertz}, and a weaker \SI{5}{\micro\second}, \SI{1}{\milli\watt} pulse of varying frequency. We use the former to simulate the heating occurring in a circuit, while the latter probes the NV centers' zero-field resonance as in a conventional ODMR experiment\cite{Schirhagl2014}. Additionally, the PL signal is collected uniquely in the presence of the microwave probing pulse, and a \SI{500}{\nano\second} time buffer separates the heating and the probing stages of the measurements to guarantee the absence of spurious frequency components from the strong heating pulse during the photon collection time bin. For each frequency of the microwave probing signal we repeat the measuring sequence until we average the signal for \SI{475}{\milli\second} and the resulting ODMR traces are compared with calibration measurements collected with the heating pulse turned off. In particular, the shift in the zero-field splitting parameter $D$ is calculated for each particle by subtracting the values obtained from a Lorentzian fit of the ODMR spectra collected with and without the heating pulse. 

From the shifts in $D$ we then estimate the values for the absolute temperature using $\frac{dD(T)}{dT}$ = \SI{100}{kHz/K}\cite{Toyli2012}, and the results are plotted in \Cref{TempMap}, superimposed to a SEM image of the mapped CPW. The discrete temperature values obtained from the 81 sensing regions (the center of which as obtained from confocal fluorescence measurements is indicated with white dots in \Cref{TempMap}) are linearly interpolated to create a continuous map. We note that enhanced spatial resolution can be obtained for this measurement with nanodiamond arrays of reduced pattern pitch and spot size. Ultimately the technique can achieve diffraction-limited resolution and be further integrated with super-resolution microscopy techniques. We confirm that the microwave signal used to probe the NV centers does not introduce observable spurious heating effects by collecting ODMR measurements at different probe powers (up to 4 times higher), which do not show any detectable shift of the ODMR resonances.

To explore our results in detail we developed a simulated temperature map of the system obtained with COMSOL Multiphysics\textsuperscript{\textregistered} (see \Cref{TempMapSim}). The model comprises both the sample (the GGG layer, the CPW, and the PDMS film) and a copper substrate on which the sample is mounted (see Supplementary Information for further details). Due to the complexity in modeling the heat transfer through wire connections to the CPW pads and to the rest of the setup, we treat the value of the heat transfer coefficient from the bottom of the copper substrate as a free parameter in the simulation to obtain the same range of temperature that we experimentally measure. Nevertheless, we obtain a value for this parameter that is physically compatible with the heat sink on which the sample holder is mounted, and we therefore expect the results of the simulation to correctly represent the temperature variations across the sample. It is important to underline that no significant differences in the temperature profile emerge when the simulation is repeated without the PDMS layer, confirming that the low thermal conductivity of PDMS minimizes spurious effects on the thermal measurements.

Comparing \Cref{TempMapSim} with \Cref{TempMap}, we see that the temperature map we obtain agrees well with the simulated results. In particular, as expected, we detect a higher temperature at the position of the signal line of the waveguide, and a shift towards lower temperatures with increasing distance from it. We also accurately image the presence of a hotter region in the inside corner where the signal line bends, which is associated to a higher current density at the inside edge of the metal. 

Using the \SI{95}{\%} confidence interval $\epsilon_{res}$ on the resonance position obtained from the ODMR data fits we can estimate the precision of our measurements defined as the temperature variation at the location of the sensor that would result in a shift of the resonance position equivalent to $\epsilon_{res}$. The average precision across the measured particle array is \SI{3.9}{} $\pm$ \SI{2.9}{\kelvin}. We note that, once a full calibration of the array is obtained (requiring the measurement of $D$ as a function of a known temperature for all the particles), the measurements can be performed at a fixed microwave probing frequency\cite{Schirhagl2014}, thereby greatly reducing the acquisition time or improving the accuracy for the same amount of signal averaging (for shot noise limited measurements). 

In conclusion, we demonstrate that, using a combination of a directed assembly chemical patterning technique and a transfer printing process, we can fabricate arrays of nanodiamonds embedded in a transferable, transparent, and flexible matrix. Through the control of the chemical patterning paramefters we show the possibility of obtaining arrays with a wide range of nanoparticles per site. In particular, we can vary from \SI{60}{\%} coverage arrays with less than two NDs per spot on average, which are ideal for single ND or single NV center applications, to \SI{\sim 100}{\%} coverage samples with up to \SI{\sim 10}{NDs} per spot that are optimal when high signals are advantageous. This transfer printing technique could also be used to create arrays with high quality, top-down fabricated nanodiamonds\cite{Andrich2014}, which would combine the portability of the PDMS system with optimal sensitivity properties.

Finally, we showcase the temperature sensing potential of the ND arrays by imaging the thermal spatial signature of an operating CPW at the micrometer scale. This capability could have a great impact for the microelectronics community, where detailed knowledge of the temperature distribution and the mapping of hot spots is crucial for the devices' reliability and lifetime. We note that it is not necessary for the target sample to support the microwave signal used to probe the NV centers. This signal can be applied using an antenna placed close to the PDMS layer or patterned on its surface opposite to the ND arrays. In addition, all-optical techniques that do not require the NV center's microwave control could also be used to perform the temperature measurements\cite{Plakhotnik2015}. When combined with a widefield imaging apparatus and ODMR measurements at a fixed probing frequency, these systems could allow for fast and sensitive mapping of large sensor arrays while maintaining microscale resolution.

\begin{acknowledgement}
The design and fabrication of the diamond nanoparticle arrays was supported by the US Department of Energy, Office of Science, Basic Energy Sciences, Materials Sciences and Engineering Division.  This work was also supported by the Army Research Office through the MURI program W911NF-14-1-0016 and U.S. Air Force Office of Scientific Research FA8650-090-D-5037.  This work made use of shared facilities supported by the NSF MRSEC Program under DMR-0820054. The authors thank A. L. Crook, P. C. Jerger, and J. C. Karsch for useful discussions.
\end{acknowledgement}



\providecommand{\latin}[1]{#1}
\makeatletter
\providecommand{\doi}
{\begingroup\let\do\@makeother\dospecials
	\catcode`\{=1 \catcode`\}=2 \doi@aux}
\providecommand{\doi@aux}[1]{\endgroup\texttt{#1}}
\makeatother
\providecommand*\mcitethebibliography{\thebibliography}
\csname @ifundefined\endcsname{endmcitethebibliography}
{\let\endmcitethebibliography\endthebibliography}{}

\end{document}